\documentclass[showpacs,amssymb,aps]{revtex4}

\usepackage{graphicx}
\begin{document}

\title{Light-front Schwinger Model at Finite Temperature}  

\author{Ashok Das and  Xingxiang Zhou}
\affiliation{Department of Physics and Astronomy,
University of Rochester,
Rochester, NY 14627-0171, USA}

\bigskip

\begin{abstract}
We study the light-front Schwinger model at finite temperature
following the recent proposal in \cite{alves}. We show that the
calculations  are
carried out efficiently by working with the full propagator for the
fermion,  which also 
avoids subtleties that arise with light-front regularizations. We
demonstrate this with the calculation of the zero temperature
anomaly. We show that temperature dependent corrections to the anomaly
vanish, consistent with the results from the calculations in the
conventional quantization. The gauge self-energy is seen to have the
expected non-analytic behavior at finite temperature, but does not
quite coincide with the conventional results. However,  the
two structures are exactly the same on-shell. We show that temperature
does  not
modify the bound state equations and that the fermion condensate has
the same behavior at finite temperature as that obtained in the
conventional quantization. 

\end{abstract}


\maketitle

\section{Introduction}

In an earlier paper \cite{alves}, it was shown that light-front field
theories \cite{brodsky} do
not admit a naive generalization to finite temperature. A proper
thermal description of such theories was proposed in ref \cite{alves}, the
meaning of which was clarified nicely by Weldon \cite{weldon}. The
calculations 
carried out in refs \cite{alves,weldon1} showed that the thermal
contributions  to the
self-energy in scalar theories (both in $\phi^{4}$ and $\phi^{3}$
theories) at one loop coincide with the result from a conventional
calculations. In this paper, we extend such an investigation to
fermionic and gauge theories. In particular, we study various
questions of interest within the context of the Schwinger model
\cite{schwinger}  at finite temperature.

As is well known, the Schwinger model which describes massless QED in
$1+1$ dimensions is exactly soluble and has been widely studied in
both conventional \cite{klaiber,elcio} as well as light-front quantization
\cite{harindranath}  at zero temperature. In this paper, we study
various questions associated with the Schwinger model in the
light-front quantization at finite temperature
\cite{kapusta,lebellac,das}. In section {\bf II},
we briefly recapitulate the finite temperature formalism proposed in
\cite{alves} for light-front quantized theories. We give explicitly
the forms of the scalar as well as the gauge boson propagators in
various gauges, both in the imaginary time as well as the real time
formalisms. The fermion case needs to be discussed carefully and we do
this separately in section {\bf III}, where we give the propagators in
both  the
imaginary time and real time formalisms. In section {\bf IV}, we
undertake a detailed study of the light-front Schwinger model at
finite temperature. While conventionally, in the light-front quantized
theories one works with only the independent fermion components, we
argue that working with the full theory can be simpler and one can
even avoid subtleties generally arising from light-front
regularizations. We show this by explicitly calculating the anomaly at
zero temperature. We then show that the finite temperature correction
to the anomaly vanishes, as is also the case in the conventional
quantization \cite{karev,das}. In the case of the light-front
Schwinger model, only one
of the fermion components thermalizes and, consequently, we show that
while the thermal self-energy for the photon has the expected
non-analytic  structure
\cite{weldon2,bedaque,hott}, it does not quite agree with the
conventional result \cite{adilson}. However, these thermal
contributions vanish on-shell in both the quantizations, once again showing
equivalence in the observable sector. We also show that the bound
state equation \cite{harindranath} remains unchanged at finite
temperature. We calculate the fermion condensate at finite
temperature \cite{jayawardene,smilga} using the method of bosonization
\cite{liguori} and show that the result in the light-front formalism
coincides with that obtained using conventional quantization. Finally,
we conclude with a brief summary in section {\bf V}.

\section{Formalism}

In this section, we will briefly recapitulate the essential results
from refs \cite{alves,weldon} and list the forms of the propagators
for scalar and gauge fields in the light-front quantization at finite
temperature. As was shown by Weldon \cite{weldon},
the proposal in \cite{alves} corresponds to choosing a coordinate system
\begin{equation}
x^{\mu}\rightarrow \bar{x}^{\mu},\qquad \mu = 0, 1,\cdots, n-1,\label{1}
\end{equation}
such that
\begin{eqnarray}
\bar{x}^{0} & = & x^{0}+x^{n-1},\nonumber\\
\bar{x}^{i} & = & x^{i},\qquad i = 1, 2,\cdots ,n-1.\label{2}
\end{eqnarray}
Such a coordinate redefinition, which does not correspond to a Lorentz
transformation, has a unit Jacobian. One can quantize the theory on
the light-front, $\bar{x}^{0}=0$, and
in this coordinate system can preserve all the simple relations of
conventional light-front quantization as well as have a thermal
description of the system with a heat bath at rest. 

Such a coordinate
system, however, has a nontrivial metric structure in the ($2\times
2$) space of ``$0$-$(n-1)$'' indices, namely,
\begin{equation}
\bar{g}^{\mu\nu} =\left(\begin{array}{rr}
0 & -1\\
-1 & -1
\end{array}\right),\qquad \bar{g}_{\mu\nu} = \left(\begin{array}{rr}
1 & -1\\
-1 & 0
\end{array}\right).\label{3}
\end{equation}
Furthermore, the energy-momentum, in the new coordinates, take the form
\begin{eqnarray}
\bar{p}_{0} & = & p_{0},\nonumber\\
\bar{p}_{n-1} & = & -p_{0} + p_{n-1},\nonumber\\
\bar{p}_{\alpha} & = & p_{\alpha},\qquad \alpha = 1,2,\cdots,
n-2,\label{4}
\end{eqnarray}
so that with (\ref{3}), the form of the Einstein relation for a
massive particle follows to be
\begin{eqnarray}
(2\bar{p}_{0}+\bar{p}_{n-1})\bar{p}_{n-1} +
  \bar{p}_{\alpha}\bar{p}_{\alpha} + m^{2} &
= & 0\nonumber\\
{\rm or,}\quad 2\bar{p}_{0}\bar{p}_{n-1} + \omega_{\bar{p}}^{2}
& = & 0,\label{5}
\end{eqnarray}
where we have defined
\begin{equation}
\omega_{\bar{p}}^{2} = \bar{p}_{i}\bar{p}_{i} + m^{2},\qquad i=1,
2,\cdots, n-1.\label{6}
\end{equation}

In this case, the density matrix for a system, interacting
with a heat bath at rest, takes the form
\begin{equation}
\rho (\beta) = e^{-\beta \bar{p}_{0}},\label{7}
\end{equation}
where $\beta = \frac{1}{T}$ in units of the Boltzmann
constant. (We note parenthetically that $\bar{p}_{0} = p_{0} =
\frac{1}{\sqrt{2}} (p^{+}+p^{-})$, which is what was used in
\cite{alves}.) 
The statistical description of quantum field theories can now be
developed in the standard manner \cite{kapusta,lebellac,das}. For
example, we list below the propagators for bosonic (scalar and gauge)
fields in both imaginary and real time (closed time path)
formalisms. The propagators for the fermions would be discussed
separately in the next section.

\subsection{Scalar Propagator}

At zero temperature, the propagator for a scalar field has the form
\begin{equation}
iG (\bar{p}) =  \frac{i}{-(2\bar{p}_{0}\bar{p}_{n-1} +
  \omega_{\bar{p}}^{2}) + i\epsilon}.\label{8}
\end{equation}
The form of the propagator, in the imaginary time formalism, now
follows from (\ref{8}) to be
\begin{equation}
G^{(\beta)} (\bar{p}) = 
\frac{1}{2\bar{p}_{0}\bar{p}_{n-1} +
  \omega_{\bar{p}}^{2}},\label{9}
\end{equation}
with $\bar{p}_{0} = \frac{2i\pi n}{\beta}$ where $n$ takes integer
values. 

In the real time formalism, on the other hand, the degrees of freedom
are known to double \cite{das}. Here, we note only the form of the
propagator in the closed time path formalism (a similar form can be
derived in a straightforward manner for thermofield dynamics).
\begin{eqnarray}
iG_{++} (\bar{p}) & = & 
\frac{i}{-(2\bar{p}_{0}\bar{p}_{n-1} + \omega_{\bar{p}}^{2}) +
i\epsilon} 
+ 2\pi n_{B}(|\bar{p}_{0}|) \delta
(2\bar{p}_{0}\bar{p}_{n-1} +
\omega_{\bar{p}}^{2}),\nonumber\\
iG_{+-} (\bar{p}) & = & 2\pi [\theta(-\bar{p}_{0}) +
n_{B}(|\bar{p}_{0}|)] \delta (2\bar{p}_{0}\bar{p}_{n-1} +
\omega_{\bar{p}}^{2}),\nonumber\\
iG_{-+} (\bar{p}) & = & 2\pi [\theta (\bar{p}_{0}) + n_{B}
(|\bar{p}_{0}|)] \delta (2\bar{p}_{0}\bar{p}_{n-1} +
\omega_{\bar{p}}^{2}),\nonumber\\
iG_{--} (\bar{p}) & = & \frac{i}{(2\bar{p}_{0}\bar{p}_{n-1}
  + \omega_{\bar{p}}^{2}) + i\epsilon} + 2\pi n_{B}(|\bar{p}_{0}|) \delta
(2\bar{p}_{0}\bar{p}_{n-1} +
\omega_{\bar{p}}^{2}).\label{10}
\end{eqnarray}
Here,
\begin{equation}
n_{B} (|\bar{p}_{0}|) = \frac{1}{e^{\beta |\bar{p}_{0}|} -
  1},\label{11}
\end{equation}
represents the bosonic distribution function.

\subsection{Gauge Boson Propagator}

The propagators for the gauge fields can also be derived in a
straightforward  manner. Without going into details, let us simply note here
that in the path integral formalism with a general covariant gauge,
the zero  temperature propagator has the form
\begin{equation}
iD_{\mu\nu} (\bar{p}) = - \frac{i}{\bar{p}^{2}} \left(\bar{g}_{\mu\nu}
- (1 - \xi) \frac{\bar{p}_{\mu}\bar{p}_{\nu}}{\bar{p}^{2}}\right),
\end{equation}
where $\xi$ represents the gauge fixing parameter (the Feynman
prescription is understood) and 
\begin{equation}
\bar{p}^{2} = - 2\bar{p}_{0}\bar{p}_{n-1} - \bar{p}_{i}\bar{p}_{i}.
\end{equation}
The finite temperature propagator, in the imaginary time formalism,
has the form
\begin{equation}
D_{\mu\nu}^{(\beta)} (\bar{p}) =
\frac{1}{\bar{p}^{2}}\left(\bar{g}^{(E)}_{\mu\nu} - (1 - \xi)
\frac{\bar{p}_{\mu}\bar{p}_{\nu}}{\bar{p}^{2}}\right),
\end{equation}
with $\bar{p}_{0} = \frac{2i\pi n}{\beta}$ and
$\bar{g}^{(E)}_{\mu\nu}$ is the properly rotated Euclidean metric.

Parenthetically, we note that, as has been already pointed out in
\cite{weldon}, going to oblique 
coordinates and rotating to imaginary time do not
commute. Nonetheless, once we are in the oblique coordinates, we can
go to the imaginary time formalism and our conventions for this are as
follows. 

\begin{equation}
\bar{x}^{\mu} = (-i\bar{x}^{0}_{(E)},\bar{x}^{i}_{(E)}).\label{a1}
\end{equation}
The energy-momentum vector, on the other hand, is rotated as
\begin{equation}
\bar{p}_{\mu} = (i\bar{p}_{0}^{(E)},\bar{p}_{i}),\label{a2}
\end{equation}
which is required from the analytic structure of the propagator (so
that we do not cross a singularity). Under the rotation in (\ref{a1}),
it is easy to check that the component of the metric (in the $0-(n-1)$
space) transforms as (the other components of the metric do not change)
\begin{equation}
\bar{g}^{\mu\nu (E)} = \left(\begin{array}{rr}
0 & -i\\
-i & -1
\end{array}\right),\qquad \bar{g}^{(E)}_{\mu\nu} =
\left(\begin{array}{rr}
-1 & i\\
i & 0
\end{array}\right).\label{a3}
\end{equation}
We note that, in an oblique coordinate system, the upper and lower
index tensors can be different, which is reflected in
(\ref{a3}). Transformation of other vectors and tensors can be
determined from the rotation in (\ref{a1}). 

In the real time formalism, the degrees of freedom will double and the
propagator will have the form (once again, the Feynman prescription is
understood as in (\ref{10}))
\begin{eqnarray}
iD_{\mu\nu, ++}^{(\beta)} (\bar{p}) & = & - \left(\bar{g}_{\mu\nu} - (1 -
\xi)\frac{\bar{p}_{\mu}\bar{p}_{\nu}}{\bar{p}^{2}}\right)
\left(\frac{i}{\bar{p}^{2}} + 2\pi n_{B}(|\bar{p}_{0}|) \delta
(\bar{p}^{2})\right),\nonumber\\
iD_{\mu\nu, +-}^{(\beta)} (\bar{p}) & = &  -\left(\bar{g}_{\mu\nu} - (1 - \xi)
\frac{\bar{p}_{\mu}\bar{p}_{\nu}}{\bar{p}^{2}}\right) 2\pi\left(\theta
(-\bar{p}_{0}) + n_{B}(|\bar{p}_{0}|)\right) \delta (\bar{p}^{2}),\nonumber\\
iD_{\mu\nu, -+}^{(\beta)} (\bar{p}) & = & -\left(\bar{g}_{\mu\nu} - (1 - \xi)
\frac{\bar{p}_{\mu}\bar{p}_{\nu}}{\bar{p}^{2}}\right) 2\pi \left(\theta
(\bar{p}_{0}) + n_{B} (|\bar{p}_{0}|)\right)\delta (\bar{p}^{2}),\nonumber\\
iD_{\mu\nu, --}^{(\beta)} (\bar{p}) & = & -\left(\bar{g}_{\mu\nu} - (1 - \xi)
\frac{\bar{p}_{\mu}\bar{p}_{\nu}}{\bar{p}^{2}}\right)
\left(-\frac{i}{\bar{p}^{2}} + 2\pi n_{B}(|\bar{p}_{0}|) \delta
(\bar{p}^{2})\right).
\end{eqnarray}

On the other hand, in a general axial gauge, the zero temperature
propagator in the path integral formalism takes the form
\begin{equation}
iD_{\mu\nu} (\bar{p}) = - \frac{i}{\bar{p}^{2}} \left(\bar{g}_{\mu\nu}
-
\frac{\bar{n}_{\mu}\bar{p}_{\nu}+\bar{n}_{\nu}\bar{p}_{\mu}}{\bar{n}\cdot
  \bar{p}} +
\frac{\bar{n}^{2}\bar{p}_{\mu}\bar{p}_{\nu}}{(\bar{n}\cdot
  \bar{p})^{2}} + \frac{\xi \bar{p}^{2}
  \bar{p}_{\mu}\bar{p}_{\nu}}{(\bar{n}\cdot \bar{p})^{2}}\right),
\end{equation}
where $\bar{n}^{\mu}$ represents an arbitrary vector (not necessarily
normalized). The finite temperature propagator, in this case, in the
imaginary time formalism takes the form
\begin{equation}
D_{\mu\nu}^{(\beta)} (\bar{p}) =
\frac{1}{\bar{p}^{2}}\left(\bar{g}^{(E)}_{\mu\nu} -
\frac{\bar{n}_{\mu}\bar{p}_{\nu}+\bar{n}_{\nu}\bar{p}_{\mu}}{\bar{n}\cdot 
  \bar{p}} +
\frac{\bar{n}^{2}\bar{p}_{\mu}\bar{p}_{\nu}}{(\bar{n}\cdot
  \bar{p})^{2}} + \frac{\xi \bar{p}^{2}
  \bar{p}_{\mu}\bar{p}_{\nu}}{(\bar{n}\cdot \bar{p})^{2}}\right),
\end{equation}
with $\bar{p}_{0} = \frac{2i\pi n}{\beta}$ and $n_{\mu}$ representing
the appropriately rotated Euclidean vectors. In the real time
formalism, the degrees double and we simply note here the $(++)$
component of the propagator (the other components can be determined
much like the forms given above)
\begin{equation}
iD_{\mu\nu, ++}^{(\beta)} (\bar{p}) = -\left(\bar{g}_{\mu\nu}
-
\frac{\bar{n}_{\mu}\bar{p}_{\nu}+\bar{n}_{\nu}\bar{p}_{\mu}}{\bar{n}\cdot
  \bar{p}} +
\frac{\bar{n}^{2}\bar{p}_{\mu}\bar{p}_{\nu}}{(\bar{n}\cdot
  \bar{p})^{2}} + \frac{\xi \bar{p}^{2}
  \bar{p}_{\mu}\bar{p}_{\nu}}{(\bar{n}\cdot
  \bar{p})^{2}}\right)\left(\frac{i}{\bar{p}^{2}} + 2\pi n_{B}
  (|\bar{p}_{0}|) \delta (\bar{p}^{2})\right).
\end{equation}
The gauge propagators in other gauges can, of course, be easily
derived, but these are the two most commonly used gauges in
light-front field theories, which is why we have listed their forms.

\section{Fermions}

In the light-front theories, the handling of the fermions is a little
tricky. This is because one of the components of the fermion field
becomes constrained. To see this, let us note that, under the
coordinate redefinition in (\ref{2}), the Dirac gamma matrices transform as
\begin{eqnarray}
\bar{\gamma}^{0} & = & \gamma^{0} + \gamma^{n-1},\nonumber\\
\bar{\gamma}^{i} & = & \gamma^{i},\qquad i = 1,2,\cdots,
n-1.\label{f1}
\end{eqnarray}
The transformed Dirac matrices satisfy the algebra
\begin{equation}
\left\{\bar{\gamma}^{\mu},\bar{\gamma}^{\nu}\right\} = 2
\bar{g}^{\mu\nu}.\label{f2}
\end{equation}
This implies, in particular, from (\ref{3}) that
\begin{eqnarray}
(\bar{\gamma}^{0})^{2} = 0,\quad (\bar{\gamma}^{n-1})^{2} & = & - 1,\nonumber\\
\bar{\gamma}^{0}\bar{\gamma}^{n-1} +
\bar{\gamma}^{n-1}\bar{\gamma}^{0} & = & -2.\label{f3}
\end{eqnarray}
With these matrices, we can define the projection operators
\begin{equation}
P^{+} = - \frac{1}{2} \bar{\gamma}^{n-1}\bar{\gamma}^{0},\quad P^{-} =
- \frac{1}{2} \bar{\gamma}^{0}\bar{\gamma}^{n-1},\quad
(P^{\pm})^{\dagger} = P^{\pm}, \quad P^{+} + P^{-}
= 1,\label{f4}
\end{equation}
which allows us to decompose the fermion field as
\begin{equation}
\psi_{\pm} = P^{\pm} \psi.\label{f5}
\end{equation}
In terms of these components, the Lagrangian density for a free
massive fermion takes the form
\begin{equation}
{\cal L} = i\psi_{+}^{\dagger}
(2\bar{\partial}_{0}+\bar{\partial}_{n-1})\psi_{+} -
i\psi_{-}^{\dagger} \bar{\partial}_{n-1}\psi_{-}
+ \psi_{+}^{\dagger} \bar{\gamma}^{n-1}
(i\bar{\gamma}^{\alpha}\bar{\partial}_{\alpha} + m)\psi_{-} -
\psi_{-}^{\dagger} \bar{\gamma}^{n-1}
(i\bar{\gamma}^{\alpha}\bar{\partial}_{\alpha} +
m)\psi_{+},\label{f6}
\end{equation}
where, as defined earlier, $\alpha = 1,2,\cdots, n-2$.

This shows that while the fermion component $\psi_{+}$ is dynamical,
the component $\psi_{-}$ is constrained and is related to the
dynamical  component
$\psi_{+}$. Because of this, the conventional practice in calculations
using light-front quantization is to eliminate the constrained
variable, which in some cases introduces additional contact
interactions into the theory for the dynamical component. From the
point of view of thermal field theory, however, we find that it is
more appropriate to work in the full Dirac space without eliminating
the dependent components. Even in the zero temperature
case, we find that this simplifies the calculations (in this case, of course,
one does not have to worry about additional contact interactions which
is certainly an advantage) as
we will demonstrate with the calculation of the anomaly in the
Schwinger model in the next section.

From the fact that $\bar{\gamma}^{0}$ is nilpotent (see (\ref{f3}))
and the forms of the projection operators in (\ref{f4}), we see that
\begin{equation}
\overline{\psi}_{+} = \psi_{+}^{\dagger} \bar{\gamma}^{0} = 0.
\end{equation}
As a result, in such a theory, it is more appropriate to define a
fermion propagator as
\begin{equation}
iS (\bar{x}-\bar{x}') = \langle T(\psi (\bar{x})\psi^{\dagger}
(\bar{x}'))\rangle,
\end{equation}
where ``$T$'' represents ordering with respect to $\bar{x}^{0}$. We
will be working in the path integral formalism, where the 
propagator would simply represent the inverse of the two point
function. In the momentum space, this then leads to the complete
propagator  for
a massive fermion at zero temperature (in the path integral
formalism) to be
\begin{equation}
iS (\bar{p}) = \frac{i}{-(2\bar{p}_{0}\bar{p}_{n-1} +
  \omega_{\bar{p}}^{2}) + i\epsilon} \left(\begin{array}{cc}
-\bar{p}_{n-1} &
  -\bar{\gamma}^{n-1}(-\bar{\gamma}^{\alpha}\bar{p}_{\alpha} + m)\\
\bar{\gamma}^{n-1} (-\bar{\gamma}^{\alpha}\bar{p}_{\alpha} + m) &
  2\bar{p}_{0}+\bar{p}_{n-1}
\end{array}\right).\label{f7}
\end{equation}
It can be easily checked that this reduces to the conventional
propagator for the dynamical independent component when restricted to
the  proper projection
\cite{alves,brodsky1}. We note here that this propagator can also be
written as
\begin{equation}
iS (\bar{p}) = \frac{i (\bar{p}\!\!\!\slash +
  m)(\bar{\gamma}^{0}-\bar{\gamma}^{n-1})}{-(2\bar{p}_{0}\bar{p}_{n-1} +
  \omega_{\bar{p}}^{2}) + i\epsilon}.\label{f7'}
\end{equation}

It now follows from (\ref{f7}) that the fermion propagator
at finite temperature  in the imaginary time formalism takes the form
\begin{equation}
S^{(\beta)} (\bar{p}) = \frac{1}{2\bar{p}_{0}\bar{p}_{n-1} +
  \omega_{\bar{p}}^{2}} \left(\begin{array}{cc}
-\bar{p}_{n-1} &
 - \bar{\gamma}^{n-1}(-\bar{\gamma}^{\alpha}\bar{p}_{\alpha} + m)\\
\bar{\gamma}^{n-1} (-\bar{\gamma}^{\alpha}\bar{p}_{\alpha} + m) &
  2\bar{p}_{0}+\bar{p}_{n-1}
\end{array}\right),\label{f8}
\end{equation}
with $\bar{p}_{0} = \frac{2i\pi (n+1)}{\beta}$.

In the real time formalism, the four components of the matrix
propagator take the forms
\begin{eqnarray}
iS_{++}^{(\beta)} (\bar{p}) & = & \left(\begin{array}{cc}
-\bar{p}_{n-1} &
 - \bar{\gamma}^{n-1}(-\bar{\gamma}^{\alpha}\bar{p}_{\alpha} + m)\\
\bar{\gamma}^{n-1} (-\bar{\gamma}^{\alpha}\bar{p}_{\alpha} + m) &
  2\bar{p}_{0}+\bar{p}_{n-1}
\end{array}\right) \left(\frac{i}{-(2\bar{p}_{0}\bar{p}_{n-1} +
  \omega_{\bar{p}}^{2}) + i\epsilon} - 2\pi n_{F} (|\bar{p}_{0}|) \delta
  (2\bar{p}_{0}\bar{p}_{n-1}+\omega_{\bar{p}}^{2})\right),\nonumber\\
\noalign{\vskip 4pt}%
iS_{+-}^{(\beta)} (\bar{p}) & = & 2\pi \left(\begin{array}{cc}
-\bar{p}_{n-1} &
 - \bar{\gamma}^{n-1}(-\bar{\gamma}^{\alpha}\bar{p}_{\alpha} + m)\\
\bar{\gamma}^{n-1} (-\bar{\gamma}^{\alpha}\bar{p}_{\alpha} + m) &
  2\bar{p}_{0}+\bar{p}_{n-1}
\end{array}\right) \left(\theta(-\bar{p}_{0}) - n_{F}
  (|\bar{p}_{0}|)\right) \delta
  (2\bar{p}_{0}\bar{p}_{n-1}+\omega_{\bar{p}}^{2}),\nonumber\\
\noalign{\vskip 4pt}%
iS_{-+}^{(\beta)} (\bar{p}) & = & 2\pi \left(\begin{array}{cc}
-\bar{p}_{n-1} &
 - \bar{\gamma}^{n-1}(-\bar{\gamma}^{\alpha}\bar{p}_{\alpha} + m)\\
\bar{\gamma}^{n-1} (-\bar{\gamma}^{\alpha}\bar{p}_{\alpha} + m) &
 2\bar{p}_{0}+\bar{p}_{n-1}
\end{array}\right)\left(\theta (\bar{p}_{0}) - n_{F}
  (|\bar{p}_{0}|)\right) \delta
  (2\bar{p}_{0}\bar{p}_{n-1}+\omega_{\bar{p}}^{2}),\nonumber\\
\noalign{\vskip 4pt}%
iS_{--}^{(\beta)} (\bar{p}) & = & \left(\begin{array}{cc}
-\bar{p}_{n-1} &
 - \bar{\gamma}^{n-1}(-\bar{\gamma}^{\alpha}\bar{p}_{\alpha} + m)\\
\bar{\gamma}^{n-1} (-\bar{\gamma}^{\alpha}\bar{p}_{\alpha} + m) &
 2\bar{p}_{0}+\bar{p}_{n-1}
\end{array}\right)\left(\frac{i}{(2\bar{p}_{0}\bar{p}_{n-1} +
  \omega_{\bar{p}}^{2}) + i\epsilon} - 2\pi n_{F} (|\bar{p}_{0}|) \delta
  (2\bar{p}_{0}\bar{p}_{n-1}+\omega_{\bar{p}}^{2})\right)\!\!,\label{f9}
\end{eqnarray}
where
\begin{equation}
n_{F} (|\bar{p}_{0}|) = \frac{1}{e^{\beta |\bar{p}_{0}|} +
  1},\label{f10}
\end{equation}
represents the fermion distribution function. We note here that the
propagators in (\ref{f9}) can be checked to reduce to the ones in
\cite{alves} when restricted to the dynamical independent components
only. However, as we will show next with the Schwinger model as an
example, in actual calculations, it may be more efficient to use the
full form of the propagator.

\section{Schwinger Model}

The Schwinger model describes quantum electrodynamics in $(1+1)$ dimensions
for massless fermions and represents an exactly soluble model. The
Lagrangian density in the new coordinates takes the form
\begin{equation}
{\cal L} = i\psi_{+}^{\dagger}(2\bar{\partial}_{0}+\bar{\partial}_{1})
\psi_{+} - i\psi_{-}^{\dagger}\bar{\partial}_{1}\psi_{-} + e
\psi^{\dagger} (\bar{\gamma}^{0} -
\bar{\gamma}^{1})\bar{\gamma}^{\mu}\psi \bar{A}_{\mu},\quad
\mu=0,1.\label{sm1} 
\end{equation}
We note that we have not separated out the interaction term into
components (although one can do so easily and the interaction is
diagonal in the Dirac space) just to demonstrate that
calculations with the full propagator may be more efficient. In
$(1+1)$ dimensions, some further simplifications occur. First of all,
in this case, we can show that, under the coordinate transformation,
the Levi-Civita tensor remains invariant (the magnitude of the
determinant of the metric is unity) and we can identify
\begin{equation}
\bar{\gamma}_{5} = -\frac{1}{2}\bar{\epsilon}_{\mu\nu}
\bar{\gamma}^{\mu}\bar{\gamma}^{\nu},\quad (\bar{\gamma}_{5})^{2} =
1,\quad  \left\{\bar{\gamma}_{5},\bar{\gamma}^{\mu}\right\} = 0.\label{sm2}
\end{equation}
It follows then, from the definitions of the projection operators in
(\ref{f4}) as well as the algebraic relations in (\ref{f3}) that
\begin{equation}
P^{\pm} = \frac{1}{2} (1 \pm \bar{\gamma}_{5}).\label{sm3}
\end{equation}
Furthermore, the usual duality relations continue to hold in this
case, namely,
\begin{equation}
\bar{\gamma}_{5}\bar{\gamma}^{\mu} = \bar{\epsilon}^{\mu\nu}
\bar{\gamma}_{\nu},\quad \bar{\gamma}^{\mu} =
\bar{\epsilon}^{\mu\nu} \bar{\gamma}_{5}\bar{\gamma}_{\nu}.\label{sm4}
\end{equation}
In this case, from the invariances of the theory as well as
(\ref{sm4}), we  can identify
\begin{equation}
\bar{\jmath}^{\mu} = \psi^{\dagger}
(\bar{\gamma}^{0}-\bar{\gamma}^{1})\bar{\gamma}^{\mu}\psi,\quad
\bar{\jmath}_{5}^{\mu} =
\psi^{\dagger}(\bar{\gamma}^{0}-\bar{\gamma}^{1})\bar{\gamma}_{5}
\bar{\gamma}^{\mu} \psi = \bar{\epsilon}^{\mu\nu}
\bar{\jmath}_{\nu}.\label{sm5}
\end{equation}
Let us next calculate the anomaly in the chiral current 
$\bar{\jmath}_{5}^{\mu}$ - both at zero temperature as well as at finite
temperature. 

\begin{figure}[h]
\begin{center}
\includegraphics[width=5cm,height=3cm]{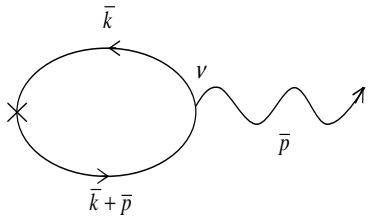}
\caption{One-loop graph contributing to the anomaly in the chiral
current. The solid lines represent fermions, the wavy line a photon
  and the cross denotes the divergence of the chiral current.}
\end{center}
\end{figure}

The calculation of the anomaly is best carried out in the real time
formalism. First, we note that, in the case of the Schwinger model, we
cannot simply take over the form of the fermion propagator from
(\ref{f9}). This is because, for a massless fermion in $1+1$ dimensions,
the non-dynamical component of the fermion is, in fact, decoupled from
the dynamical one as is obvious from (\ref{sm1}). As a result, even at
zero temperature, the form of the propagator has the form (with the
Feynman prescription understood, see also
(\ref{f7}))
\begin{eqnarray}
iS (\bar{p}) & = &
\frac{i}{-(2\bar{p}_{0}+\bar{p}_{1})\bar{p}_{1}}\left(\begin{array}{cc} 
-\bar{p}_{1} & 0\\
0 & 2\bar{p}_{0}+\bar{p}_{1}
\end{array}\right) = \left(\begin{array}{cc}
\frac{i}{2\bar{p}_{0}+\bar{p}_{1}} & 0\\
0 & - \frac{i}{\bar{p}_{1}}
\end{array}\right)\nonumber\\
\noalign{\vskip 4pt}%
 & = & \frac{i}{2\bar{p}_{0}+\bar{p}_{1}} P^{+} -
 \frac{i}{\bar{p}_{1}} P^{-}.\label{sm6}
\end{eqnarray}
As in (\ref{f7'}), we note here that this propagator can also be
written as
\begin{equation}
iS (\bar{p}) = \frac{i \bar{\gamma}^{\mu}\bar{p}_{\mu}
(\bar{\gamma}^{0}-\bar{\gamma}^{1})}{-(2\bar{p}_{0}+\bar{p}_{1})\bar{p}_{1}},
\label{sm6'}
\end{equation}
which reflects the propagator relations between the conventional
quantization and the light-front quantization from the point of view
of a coordinate transformation.

The zero temperature anomaly can be calculated using the component
fields (which is what is conventionally done in the light-front
studies). However, we wish to point out that it is equally convenient
to carry out the light-front calculations using the full propagator
and the complete vertex of the theory. Both lead to the same result,
however, using the full fermion propagator and the complete vertex,
one can  avoid
subtleties arising from regularization. To demonstrate this, let us
calculate the zero temperature anomaly in the Schwinger model using
the full propagator in (\ref{sm6'}) and the complete vertex in
(\ref{sm1}). We note from Fig. 1 that the zero temperature amplitude has
the form
\begin{eqnarray}
 & = & -e^{2} \int\frac{d^{2}\bar{k}}{(2\pi)^{2}}\,{\rm
Tr}\,(\bar{\gamma}^{0}-\bar{\gamma}^{1})
\bar{\gamma}_{5}\bar{p}\!\!\!\slash S(\bar{k})
(\bar{\gamma}^{0}-\bar{\gamma}^{1})\bar{\gamma}^{\nu}
S(\bar{k}+\bar{p})\nonumber\\
 & = & -e^{2} \int \frac{d^{2}\bar{k}}{(2\pi)^{2}}\,{\rm
Tr}\,\bar{\gamma}_{5}
\bar{p}\!\!\!\slash\bar{k}\!\!\!\slash\bar{\gamma}^{\nu}(\bar{k}\!\!\!\slash
+
\bar{p}\!\!\!\slash)\,\frac{1}{\bar{k}^{2}(\bar{k}+\bar{p})^{2}}.\label{sm7}
\end{eqnarray}
Here, we have used the form of the full propagator in (\ref{sm6'}),
the notation
\begin{equation}
\bar{k}^{2} = - (2\bar{k}_{0}+\bar{k}_{1})\bar{k}_{1},\label{sm8}
\end{equation}
as well as the identity
\begin{equation}
(\bar{\gamma}^{0}-\bar{\gamma}^{1})^{2} = 1.\label{sm9}
\end{equation}

There are several things to note from Eq. (\ref{sm7}). First, the form
of the integrand is the same as would be obtained in the conventional
quantization except for barred quantities. However, since scalar
quantities are unchanged under the coordinate transformation (\ref{2})
and vectors transform in a simple manner, we expect the results to be
quite similar to the standard result. In fact, the fermion trace leads
to the same result in the barred variables. Normally, the light-front
integrals have to be treated with care, but with the full propagator,
we note that we can make a change of variables of integration
(basically, the inverse redefinition of (\ref{2}))
\begin{equation}
\bar{k}_{0} = k_{0},\quad \bar{k}_{3} = - k_{0} + k_{3},\label{sm10}
\end{equation}
which allows us to use standard dimensional regularization (in this
case, of course, the result turns out to be finite because of gauge
invariance) leading to the value of the amplitude 
\begin{equation}
- \frac{e^{2}}{\pi} \bar{\epsilon}^{\mu\nu} \bar{p}_{\mu},
\end{equation}
which determines that the anomaly in the chiral current, at zero
temperature, is given by
\begin{equation}
\bar{\partial}_{\mu} \bar{\jmath}_{5}^{\mu} = - \frac{e^{2}}{2\pi}
\bar{\epsilon}^{\mu\nu} \bar{F}_{\mu\nu}.\label{sm11}
\end{equation}

To calculate the thermal correction to the anomaly, we note that in
the imaginary time formalism, the fermion propagator has the
form (see Eq. (\ref{sm6}))
\begin{equation}
S^{(\beta)} (\bar{p}) =  -\frac{1}{2\bar{p}_{0}+\bar{p}_{1}} P^{+} +
\frac{1}{\bar{p}_{1}} P^{-},\label{sm12}
\end{equation}
with $\bar{p}_{0} = \frac{(2n+1)i\pi}{\beta}$. For the calculation of
the anomaly, however, the real time formalism is more suitable and the
propagator, in this case, is given by (we give only the $++$ component
which is relevant)
\begin{equation}
iS_{++}^{(\beta)} (\bar{p}) = \bar{\gamma}^{\mu}\bar{p}_{\mu}
(\bar{\gamma}^{0}-\bar{\gamma}^{1})\left(
\frac{i}{-(2\bar{p}_{0}+\bar{p}_{1})\bar{p}_{1} + i\epsilon} - P^{+} 2\pi
n_{F}(|\bar{p}_{0}|) \delta
((2\bar{p}_{0}+\bar{p}_{1})\bar{p}_{1})\right),\label{sm13}
\end{equation}
where $n_{F}$ represents the fermion distribution function defined in
(\ref{f10}). 
The interesting thing to note from (\ref{sm13}) is the presence of the
projection operator $P^{+}$ in the thermal term. This simply reflects
the fact that the fermion component $\psi_{-}$ is nondynamical and, as
a consequence, does not thermalize. This is also reflected in the form
of the propagator (\ref{sm12}) in the imaginary time formalism where
the component involving $P^{-}$ has no $\bar{p}_{0}$ dependence and,
consequently, does not have any temperature dependence. We would like
to emphasize that the propagator in (\ref{sm13}) also results if we
start with a massive fermion propagator as in (\ref{f9}) (for $1+1$
dimensions) and take the limit $m\rightarrow 0$.

With the form of the propagator in (\ref{sm13}), we can now calculate
the temperature dependence of the amplitude in Fig. 1. Once again, the
calculations can be done in components or with the full propagator and
both yield the same result. If we take the full propagator and the
vertex, the temperature dependent part of the amplitude can be
calculated very easily. The terms linear in the fermion distribution
function (with a little algebra) take the form
\begin{equation}
\sim \int
\frac{d^{2}\bar{k}}{(2\pi)^{2}}\,\bar{\epsilon}^{\mu\nu}\bar{k}_{\mu}\,
n_{F}(|\bar{k}_{0}|) \delta (\bar{k}^{2}) = 0.\label{sm14}
\end{equation}
Similarly, the terms in the amplitude quadratic in the fermion
distribution function give (even before doing the Dirac trace)
\begin{equation}
\sim \int \frac{d^{2}\bar{k}}{(2\pi)^{2}}\,{\rm Tr}\, (1
+\bar{\gamma}_{5})\bar{k}\!\!\!\slash \bar{\gamma}^{\nu}
(\bar{p}^{2}+2\bar{p}\cdot \bar{k})
n_{F}(|\bar{k}_{0}|)n_{F}(|\bar{k}_{0}+\bar{p}_{0}|) \delta
(\bar{k}^{2})\delta ((\bar{k}+\bar{p})^{2}) = 0.\label{sm15}
\end{equation}
This shows that the anomaly is unchanged by temperature corrections
which is, of course, well known in the conventional quantization
\cite{das,karev,itoyama}, but
holds true also in the light-front quantization. Since the chiral
anomaly is directly related to the mass of the photon in the Schwinger
model, this also implies that the photon mass is unchanged by the
temperature corrections.

Let us next calculate the temperature dependent correction to the
self-energy of the photon. The photon self-energy is a second rank
symmetric tensor and it is easy to see from the form of the amplitude
that 
\begin{equation}
\bar{\Pi}^{00 (\beta)} (\bar{p}) = 2 \bar{\Pi}^{01 (\beta)} (\bar{p}) = 2
\bar{\Pi}^{10 (\beta)} (\bar{p}) = 4 \bar{\Pi}^{11 (\beta)} (\bar{p}),
\label{sm16}
\end{equation}
so that only one independent component needs to be calculated. The
calculation is straightforward and leads to
\begin{equation}
\bar{\Pi}^{01 (\beta)} (\bar{p}) = - e^{2} \delta (2\bar{p}_{0}+\bar{p}_{1})
\int d\bar{k}_{1}\,{\rm sgn} (\bar{k}_{1}) {\rm sgn}
(\bar{k}_{1}+\bar{p}_{1}) \left(n_{F}(\frac{|\bar{k}_{1}|}{2}) +
n_{F}(\frac{|\bar{k}_{1}+\bar{p}_{1}|}{2}) - 2
n_{F}(\frac{|\bar{k}_{1}|}{2})n_{F}(\frac{|\bar{k}_{1}+\bar{p}_{1}|}{2})
\right).\label{sm17}
\end{equation}
There are several things to note from
Eqs. (\ref{sm16})-(\ref{sm17}). First of all, it is clear that the
self-energy is gauge invariant (transverse). Second taking the dual
(in one  of the
indices) and contracting with the external momentum gives zero, which
shows  again that the anomaly has no temperature
dependent contribution. The presence of the delta function structure
in the amplitude is a reflection of the non-analyticity in amplitudes
at finite temperature and the amplitude in (\ref{sm17}) reflects the
structure found in the conventionally quantized theory
\cite{adilson}. However, there is a difference in the sense that the
amplitude in Eq. (\ref{sm17}) shows only one delta function structure
whereas in the conventionally quantized theory, there are two
independent delta function structures present. This difference can be
traced back to the fact that in the light-front quantization of the
Schwinger model, only one of the fermion components thermalizes (which
is how one delta function structure arises and which also reflects the
fact that light-front quantization inherently breaks parity invariance
\cite{dharam}). Thus, it would seem that
there is finally a difference between the light-front and
conventionally quantized theories. Let us recall, however, that the
photon is massive in the Schwinger model whereas the thermal
self-energies in (\ref{sm16}), (\ref{sm17}) as well as those in
\cite{adilson} contribute nontrivially only when
$2\bar{p}_{0}+ \bar{p}_{1}=0$ (or $p_{0}\pm p_{1}=0$ in
\cite{adilson}).  Consequently, for a massive photon on-shell,
the thermal self-energies vanish in both conventional as well as
light-front quantizations. On the other hand, they are non-vanishing
and distinct in the two quantizations away from the physical
mass-shell. This calculation can be easily generalized to the thermal
$n$-point amplitudes for the photon completely along the lines
discussed in \cite{adilson}. Without going into technical details, we
simply summarize our result here. The non-vanishing components of the
thermal $n$-point amplitude have an identical structure to that in
\cite{adilson} except that we find only a single product of delta
functions of the kind in (\ref{sm17}) (which again reflects that only
one fermion component thermalizes). Once again, this shows that these
thermal amplitudes vanish on-shell for a massive photon as is the case
in \cite{adilson}, but off-shell, the two results are quite distinct.

An important aspect of the light-front quantization is that it allows for
a simpler description of questions involving bound states. From our
discussion above, since the on-shell thermal self-energy for the
photon vanishes, the equation for the bound state of fermions (and,
therefore, the solution) should not change at finite temperature. This
can also be seen quantitatively as follows. In the axial gauge (which
is conventionally used in the study of bound states in this problem),
$\bar{A}_{1}=0$, the photon equation becomes a constraint. In fact,
the Lagrangian density for a massive fermion (the mass parameter can be
taken to zero) interacting with an electromagnetic potential in the
axial gauge takes the form
\begin{equation}
{\cal L} = \frac{1}{2} (\bar{\partial}_{1}\bar{A}_{0})^{2} +
i\psi_{+}^{\dagger}(2\bar{\partial}_{0}+\bar{\partial}_{1})\psi_{+} -
i\psi_{-}^{\dagger} \bar{\partial}_{1}\psi_{-} +
m\psi_{+}^{\dagger}\bar{\gamma}^{1}\psi_{-} -
m\psi_{-}^{\dagger}\bar{\gamma}^{1}\psi_{+} + 2e
\psi_{+}^{\dagger}\psi_{+}\bar{A}_{0},\label{sm18}
\end{equation}
leading to the equation for the photon of the form
\begin{equation}
\bar{A}_{0} =
\frac{2e}{(\bar{\partial}_{1})^{2}}\,\psi_{+}^{\dagger}\psi_{+}.\label{sm19}
\end{equation}
Eliminating this constraint, the Hamiltonian takes the form
\begin{equation}
H = \int
d\bar{x}^{1}\left[-i\psi_{+}^{\dagger}\bar{\partial}_{1}\psi_{+}
+im^{2}\psi_{+}^{\dagger}\frac{1}{\bar{\partial}_{1}}\psi_{+} -
2e^{2}\psi_{+}^{\dagger}\psi_{+} \frac{1}{(\bar{\partial}_{1})^{2}}
\psi_{+}^{\dagger}\psi_{+}\right].\label{sm20}
\end{equation}
The self-energy term depends on $\frac{1}{(\bar{\partial}_{1})^{2}}$,
which does not change at finite temperature. As a result, the bound
state equation as well as the solution remains unchanged at finite
temperature. 

Another interesting quantity that can be calculated in this model is
the fermion condensate. There are various ways of calculating
this at finite temperature\cite{jayawardene,smilga}. However, we
follow, for simplicity, the  method in
\cite{smilga} which uses bosonization and is relatively
straightforward. The bosonized version of the Schwinger model
describes a free, massive scalar field
\begin{equation}
{\cal L} = \frac{1}{2} \bar{\partial}_{\mu}\phi
\bar{\partial}^{\mu}\phi - \frac{m_{\rm ph}^{2}}{2} \phi^{2},\label{sm
21}
\end{equation}
where
\begin{equation}
m_{\rm ph} = \frac{e}{\sqrt{\pi}}.\label{sm22}
\end{equation}
The correspondence between the bosonic and the fermionic degrees of
freedom, among other things, leads to the identification
\begin{equation}
\bar{\psi}\psi = - \frac{m_{\rm ph}e^{\gamma}}{2\pi} :\cos
\sqrt{4\pi}\phi :\,.\label{sm23}
\end{equation}
Here $\gamma$ represents Euler's constant and the ``colons'' stand for
normal ordering with respect to the scalar annihilation and creation
operators. It is straightforward to calculate from this the value of
the fermion condensate at zero temperature,
\begin{equation}
\langle \bar{\psi}\psi\rangle_{T=0} = - \frac{m_{\rm
ph}e^{\gamma}}{2\pi},\label{sm24}
\end{equation}
since the normal ordered fields lead to trivial vacuum expectation
values at zero temperature. At finite temperature, on the other hand,
the condensate has the form
\begin{equation}
\langle \bar{\psi}\psi\rangle_{T} = \langle
\bar{\psi}\psi\rangle_{T=0}\,e^{-2i\pi (G^{(\beta)} (0) - G
(0))}.\label{sm25}
\end{equation}
Using the representation for the scalar propagator in (\ref{10}) (in
$1+1$ dimensions and using only the $++$ component), it is easy to see that
\begin{equation}
iG^{(\beta)} (0) - i G (0) = \frac{1}{2\pi} \int_{0}^{\infty}
\frac{d\bar{k}_{1}}{\bar{k}_{1}}\,\frac{1}{e^{\beta(\bar{k}_{1} +
\frac{m_{\rm ph}^{2}}{\bar{k}_{1}})} - 1}.\label{sm26}
\end{equation}

The integral in (\ref{sm26}) cannot be evaluated in closed form in
general.  However,
for low temperatures (large $\beta$), it has the form
\cite{gradshteyn}
\begin{eqnarray}
iG^{(\beta)} (0) - i G (0) & \approx & \frac{1}{2\pi}
\int_{0}^{\infty}
\frac{d\bar{k}_{1}}{\bar{k}_{1}}\,e^{-\beta(\bar{k}_{1} + \frac{m_{\rm
ph}^{2}}{\bar{k}_{1}})} = \frac{1}{\pi} K_{0} (\beta m_{\rm ph})\nonumber\\
 & \rightarrow & \sqrt{\frac{T}{2\pi
m_{\rm ph}}}\,e^{-\frac{m_{\rm ph}}{T}}.\label{sm27}
\end{eqnarray}
On the other hand, at high temperatures (small $\beta$), we have
\cite{gradshteyn} 
\begin{eqnarray}
iG^{(\beta)} (0) - i G (0) & = & \frac{1}{2\pi} \sum_{n=1}^{\infty}
\int_{0}^{\infty}
\frac{d\bar{k}_{1}}{\bar{k}_{1}}\,e^{-n\beta(\bar{k}_{1} +
\frac{m_{\rm ph}^{2}}{\bar{k}_{1}})} = \frac{1}{\pi}
\sum_{n=1}^{\infty} K_{0} (n\beta m_{\rm ph})\nonumber\\
& \rightarrow & \frac{T}{2m_{\rm ph}} + \frac{1}{2\pi} \ln
\frac{m_{\rm ph}}{4\pi T} + \frac{\gamma}{2\pi}.\label{sm28}
\end{eqnarray}
Using Eq. (\ref{sm27}) in (\ref{sm25}), we obtain
the value of the condensate at low temperatures to be
\begin{equation}
\langle\bar{\psi}\psi\rangle_{T} \approx
\langle\bar{\psi}\psi\rangle_{T=0} \left(1-\sqrt{\frac{2\pi T}{m_{\rm
ph}}} e^{-\frac{m_{\rm ph}}{T}}\right),\label{sm29}
\end{equation}
whereas Eq. (\ref{sm28}) leads to the high temperature value of the
condensate as
\begin{equation}
\langle\bar{\psi}\psi\rangle_{T} \approx - 2T e^{-\frac{\pi
T}{m_{\rm ph}}}.\label{sm30}
\end{equation}
These are precisely the values of the condensates obtained earlier
using the conventional quantization \cite{jayawardene,smilga} and we
see once  again that the results
in the two quantizations coincide even at finite temperature.

\section{Conclusion}
In this paper, we have studied the light-front Schwinger model in
detail following the recent proposal \cite{alves}. We have shown, with
the calculation of the anomaly at zero temperature, that it may be
more efficient to calculate with the full theory when fermions are
involved. We have shown that the thermal corrections to the anomaly
vanish, consistent with the expectation from the calculations with the
conventional quantization. The thermal photon self-energy is shown to
have the expected non-analytic behavior, but coincides with the result
from the conventional quantization only on-shell. We have shown that
the bound state equations are unchanged at non zero temperature and
that the fermion condensate has the same value at finite temperature
as in  conventional quantization. In fact, if light-front quantization
is viewed as quantization in a general coordinate system
\cite{weldon},  the physical S-matrix elements will be naively
expected to be the same in both light-front as well as conventional
quantizations. At zero temperature, particularly, such an equivalence
in the physical sector, even though expected \cite{chang}, is hard to
prove rigorously owing to subtleties involving regularization of
ultraviolet divergences \cite{brodsky2}. However, the thermal
contributions are free from ultraviolet divergences and, consequently,
one may expect equivalence of physical thermal amplitudes in the two
quantizations. Our calculations, in the Schwinger model, explicitly
exhibit this feature in this model and furthermore show that off-shell
Greens functions in the two quantizations need not be the same. 

\vskip 1cm

\noindent{\bf Acknowledgment:}

One of us (AD) would like to thank Prof. J. Frenkel for helpful
discussions. This work was supported in part by US DOE Grant number DE-FG
02-91ER40685.

\vskip .5cm

\noindent{\em Note added:} In a later paper \cite{new}, complete
thermal equivalence between conventional quantization and light-front
quantization is claimed where the general proof is based on formal
arguments.  In this paper, on the other hand, we have explicitly 
evaluated the thermal amplitudes in a given theory, namely, the
Schwinger model and our calculation
shows that the off-shell thermal Green's functions, in this theory, are
different in the two quantizations. It is quite likely, therefore,
that some of the assumptions that go into the general proof are
violated in this model, as our calculation seems to suggest.

\end{document}